\documentclass[conference]{IEEEtran}
\IEEEoverridecommandlockouts
\usepackage{cite}
\usepackage{amsmath,amssymb,amsfonts}
\usepackage{algorithmic}
\usepackage{graphicx}
\usepackage{textcomp}
\usepackage{xcolor}
\usepackage{multirow}
\usepackage{array}
\usepackage{dblfloatfix}
\usepackage{soul}
\usepackage{array, makecell}
\usepackage{wrapfig}
\usepackage{xcolor}
\usepackage{hyperref}

\def\BibTeX{{\rm B\kern-.05em{\sc i\kern-.025em b}\kern-.08em
    T\kern-.1667em\lower.7ex\hbox{E}\kern-.125emX}}
\begin{document}

\title{Multi-view Representation Learning from Malware to Defend Against Adversarial Variants\\


\thanks{*: Corresponding author

Acknowledgments: This material is based upon work supported by the National Science Foundation (NSF) under Secure and Trustworthy Cyberspace (1936370)
, Cybersecurity Innovation for Cyberinfrastructure (1917117)
, and Cybersecurity Scholarship-for-Service (1921485)
programs.}
}

\makeatletter
\newcommand{\linebreakand}{%
  \end{@IEEEauthorhalign}
  \hfill\mbox{}\par
  \mbox{}\hfill\begin{@IEEEauthorhalign}
}
\makeatother

\author{

\IEEEauthorblockN{James Lee Hu*}
\IEEEauthorblockA{\textit{Department of Management Information Systems} \\
\textit{University of Arizona}\\
Tucson, USA \\
jameshu@arizona.edu}

\and

\IEEEauthorblockN{Mohammadreza Ebrahimi*}
\IEEEauthorblockA{\textit{School of Information Systems and Management} \\
\textit{University of South Florida}\\
Tampa, USA \\
ebrahimim@usf.edu}

\linebreakand

\IEEEauthorblockN{Weifeng Li}
\IEEEauthorblockA{\textit{Department of Management Information Systems} \\
\textit{University of Georgia}\\
Athens, USA \\
weifeng.li@uga.edu}

\and

\IEEEauthorblockN{Xin Li*}
\IEEEauthorblockA{\textit{Department of Computer Science} \\
\textit{University of Arizona}\\
Tucson, USA \\
xinli2@arizona.edu}

\linebreakand

\IEEEauthorblockN{Hsinchun Chen}
\IEEEauthorblockA{\textit{Department of Management Information Systems} \\
\textit{University of Arizona}\\
Tucson, USA \\
hsinchun@arizona.edu}
}

\maketitle
\thispagestyle{plain}
\pagestyle{plain}

\begin{abstract}
Deep learning-based adversarial malware detectors have yielded promising results in detecting never-before-seen malware executables without relying on expensive dynamic behavior analysis and sandbox. Despite their abilities, these detectors have been shown to be vulnerable to adversarial malware variants - meticulously modified, functionality-preserving versions of original malware executables generated by machine learning. Due to the nature of these adversarial modifications, these adversarial methods often use a \textit{single view} of malware executables (i.e., the binary/hexadecimal view) to generate adversarial malware variants. This provides an opportunity for the defenders (i.e., malware detectors) to detect the adversarial variants by utilizing more than one view of a malware file (e.g., source code view in addition to the binary view). The rationale behind this idea is that while the adversary focuses on the binary view, certain characteristics of the malware file in the source code view remain untouched which leads to the detection of the adversarial malware variants. To capitalize on this opportunity, we propose Adversarially Robust Multiview Malware Defense (ARMD), a novel multi-view learning framework to improve the robustness of DL-based malware detectors against adversarial variants. Our experiments on three renowned open-source deep learning-based malware detectors across six common malware categories show that ARMD is able to improve the adversarial robustness by up to seven times on these malware detectors.

\end{abstract}

\begin{table*} [b!]
\centering
\setlength{\abovecaptionskip}{0pt}
\setlength{\belowcaptionskip}{-10mm}
\begin{center}
\vspace{-6mm}
\caption{Selected Significant Prior Research on AMG Append Attacks against Malware Detectors}
\begin{tabular}{
|c<{\centering}
|c<{\centering}
|c<{\centering}
|c<{\centering}
|c<{\centering}|}

\hline

\textbf{Year}&\textbf{Author(s)}&\textbf{Data Source}&\textbf{Attack Method}&\textbf{View}\\

\hline

2021 & Ebrahimi et al. \cite{ebrahimi2021binary} & VirusTotal & Deep RL & Binary \& Source Code\\

\hline

2021 & Demetrio et al. \cite{demetrio2021functionality} & VirusTotal & Genetic programming & Binary\\

\hline

2021 & Hu et al. \cite{hu2021single} & VirusTotal & GPT2 & Binary\\

\hline

2020 & Ebrahimi et al. \cite{ebrahimi2020binary} & VirusTotal & Generative RNN & Binary\\

\hline

2019 & Castro et al. \cite{castro2019armed} & VirusTotal & Random perturbations & Binary\\

\hline

2019 & Chen et al. \cite{chen2019adversarial} & VirusShare, Malwarebenchmark & Enhanced random perturbations & Binary\\

\hline

2019 & Dey et al. \cite{dey2019evadepdf} & Contagio PDF malware dump & Genetic programming & Binary\\

\hline

2019 & Fang et al. \cite{fang2019evading} & VirusTotal & Deep RL & Binary\\

\hline

2019 & Park et al. \cite{park2019generation} & Malmig \& MMBig & Dynamic programming & Source Code\\

\hline

2019 & Rosenberg et al. \cite{rosenberg2019defense} & VirusTotal & GAN & API Call\\

\hline

2019 & Suciu et al. \cite{suciu2019exploring} & VirusTotal, Reversing Labs, FireEye & Append attack & Binary\\

\hline

2018 & Anderson et al. \cite{anderson2018learning} & VirusTotal & Deep RL & Binary \& Source Code\\

\hline

\end{tabular}
\vspace{2mm}

{\centering \textbf{Note:} RNN: Recurrent Neural Network; NN: Neural Network; GAN: Generative Adversarial Network; RL: Reinforcement Learning\par} 

\label{lit_overview}
\end{center}
\vspace{-7mm}
\end{table*}

\begin{IEEEkeywords}
Multi-View Learning, Adversarial Machine Learning, Adversarial Malware Variants, Deep Learning-based Malware Detectors, Adversarial Robustness 
\end{IEEEkeywords}
\section{Introduction}

Recent studies have shown deep Learning (DL)-based malware detectors are susceptible to attacks from Adversarial Malware Generation (AMG) techniques \cite{ebrahimi2021binary}\cite{hu2021single}\cite{ebrahimi2020binary}. These AMG methods automatically generate adversarial malware variants to evade a targeted malware detector. The generated malware samples can be used to improve a malware detector's robustness against such attacks \cite{demetrio2021functionality,ebrahimi2020binary}. Most AMG methods employ additive modifications via injecting bytes to the malware binary to generate functionality-preserving evasive variants \cite{suciu2019exploring}. However, these AMG methods are often limited to operating on the binary view of a malware sample and do not account for other views representing malware samples (e.g., a malware's source code). Thus, the multi-view (MV) nature of malware could be leveraged to improve detector robustness against these adversarial variants. As such, we expect MV learning to boost adversarial robustness.

MV learning refers to a branch of machine learning models that processes multiple distinct representations from the same instance of input data \cite{sun2013survey}. When applied to malware detection, MV learning has been shown to significantly improve malware detection accuracy \cite{appice2020clustering}\cite{narayanan2018multi}. These models often leverage a mechanism to combine different views into a single representation, known as a fusion mechanism \cite{li2018survey}. However, the impact of MV learning, and its different fusion mechanisms, on detector models' adversarial robustness is unclear. We hypothesize that MV learning can detect adversarial variants evasive to single-view detectors by extracting features from malware views untouched by AMG methods. Thus, in this study, we propose Adversarially Robust Multiview Malware Defense (ARMD), an MV learning framework to improve the robustness of DL-based malware detectors against adversarial variants. 


In the remainder of this manuscript, first, we review AMG, DL-based malware detectors, MV learning, fusion mechanisms, and highway layers. Subsequently, we detail the components of our proposed framework and its contribution. We then conduct several experiments to evaluate the performance of ARMD. Lastly, we highlight promising future directions. 

\section{Literature Review}
Five areas of research are examined. First, we review extant AMG studies as the overarching area for our study. Second, we examine DL-based Malware Detectors as an effective type of AI model to detect malicious samples. Third, we review MV Learning as a potential way to boost a DL-based detector's adversarial robustness. Fourth, we investigate Fusion Mechanisms to determine their impact on an MV Learning model's adversarial robustness. Lastly, we review Highway Layers as a potential remedy for the shortcomings of existing fusion mechanisms regarding adversarial robustness. 
\subsection{Adversarial Malware Generation (AMG)}

AMG aims to perturb malware samples and generate variants that evade malware detectors. Among the prevailing AMG methods, append attacks (considered as additive modifications) are the most practical due to their high chance of preserving the functionality of the original malware executable \cite{suciu2019exploring}. We  summarize  selected  significant append-based prior work based on their data source, attack method used, and the view(s) of the malware sample they operate under in Table \ref{lit_overview}.

Three major observation are made from Table \ref{lit_overview}. First, the majority of studies use VirusTotal, an open-source online malware database, as a source of their malware samples \cite{ebrahimi2021binary}\cite{demetrio2021functionality}\cite{hu2021single}\cite{ebrahimi2020binary}\cite{castro2019armed}\cite{chen2019adversarial}\cite{fang2019evading}\cite{rosenberg2019defense}\cite{suciu2019exploring}\cite{anderson2018learning}. Second, regarding selected attack methods, a few notable attack methods include simple append attack \cite{suciu2019exploring}, attacking using randomly generated perturbation \cite{castro2019armed}, and attacking using specific perturbations that lowers a malware detector's score \cite{chen2019adversarial}. More advanced methods incorporate machine learning techniques (Genetic Programming \cite{demetrio2021functionality} \cite{dey2019evadepdf}, Gradient Descent \cite{castro2019armed}, and Dynamic Programming \cite{park2019generation}) and implement advanced DL-based techniques (Generative Adversarial Networks \cite{rosenberg2019defense}, Deep Reinforcement Learning \cite{ebrahimi2021binary}\cite{fang2019evading}\cite{anderson2018learning}, and Generative Recurrent Neural Networks \cite{ebrahimi2020binary}\cite{hu2021single}). Third, and most importantly, most AMG methods only operate within a single view of the malware. Many of these AMG methods operate in the binary view \cite{demetrio2021functionality}\cite{hu2021single}\cite{ebrahimi2020binary}\cite{castro2019armed}\cite{chen2019adversarial}\cite{dey2019evadepdf}\cite{fang2019evading}\cite{suciu2019exploring}. A few studies delved into AMG attacks on other views (e.g., the source code view \cite{park2019generation} and API call view \cite{park2019generation}). The main exceptions are two Deep RL-based AMG studies \cite{ebrahimi2021binary}\cite{anderson2018learning}. These two studies include multiple different perturbations in their RL action space, a few of which results in a simultaneous binary and source code edit. Overall, we observe that most AMG studies only operate within a single view of the malware. As such, when attacking an MV malware detector, these AMG methods are expected to be rendered ineffective due to their perturbations only affecting certain parts of the malware detector's input. 

\subsection{Deep Learning-based (DL-based) Malware Detectors}

\begin{figure}[!h]
\centering
    \vspace{-9pt}
        \includegraphics[width=0.21\textwidth]{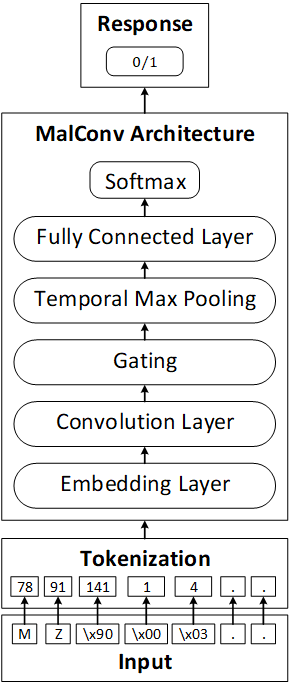}
    \vspace{-9pt}
    \caption{MalConv Architecture}
    \label{malconv}
\vspace{-10pt}
\end{figure}

DL-based malware detectors have shown high performance in malware categorization \cite{cakir2018malware}\cite{kalash2018malware}\cite{raff2018malware}. One such well-known detector is MalConv, a widely-used open-source DL-based malware detector operating only in the binary view of the malware \cite{raff2018malware}. MalConv's architecture utilizes a Convolutional Neural Network (CNN) architecture to directly process a malware sample's byte data (Figure \ref{malconv}). Specifically, MalConv takes the first 2 million bytes of a malware executable as input, passing the data through its tokenization and embedding for data pre-processing. Subsequently, the information is passed through a convolution and a max pooling layer to extract features from the malware binary regardless of location. A fully connected layer then interprets extracted features and performs classification. Lastly, a softmax layer returns a score mapped to a final result from {0,1} indicating benign or malicious. Due to its widespread popularity in malware academia as both a reference and a benchmark method, MalConv is considered representative of most open-source DL-based malware detectors \cite{ebrahimi2021binary}\cite{ebrahimi2020binary}\cite{hu2021single}\cite{fleshman2018non}\cite{krvcal2018deep}\cite{suciu2019exploring}. 

\subsection{Multi-View (MV) Learning}
MV learning refers to learning from multiple distinct representations of the same instance of data \cite{sun2013survey}. It has emerged as a viable way to enhance deep learning performance \cite{appice2020clustering}. MV malware detectors can extract salient features available in certain views, allowing them to learn a more accurate representation of the malware data \cite{narayanan2018multi}. As such, MV learning has yielded promising improvements in advanced malware classification tasks \cite{narayanan2018multi}\cite{appice2020clustering}. MV malware detectors require adversaries to attack multiple views to achieve performance comparable to attacks against single-view models \cite{sun2021adversarial}. Since recent AMG techniques focus on single-view operations, We expect that MV learning can potentially enhance a malware detector’s adversarial robustness.  

\subsection{Fusion Mechanisms}
MV learning models combine different views into a single representation for processing. Most models achieve this through a fusion mechanism \cite{li2018survey}. Advanced fusion mechanisms can also control the flow of information during fusion, improving performance.

Two dominant fusion mechanisms exist in literature: concatenation-based and attention-based fusion. Concatenation is the most basic fusion mechanism and simply combines views additively without further processing \cite{gao2020survey}. Attention-based fusion utilizes an attention mechanism to determine the importance of different areas of the input as it combines views (Figure \ref{attnFusion}) \cite{jetley2018learn}.

\begin{figure}[!h]
\vspace{-2mm}
    \centering
    \includegraphics[width=0.38\textwidth]{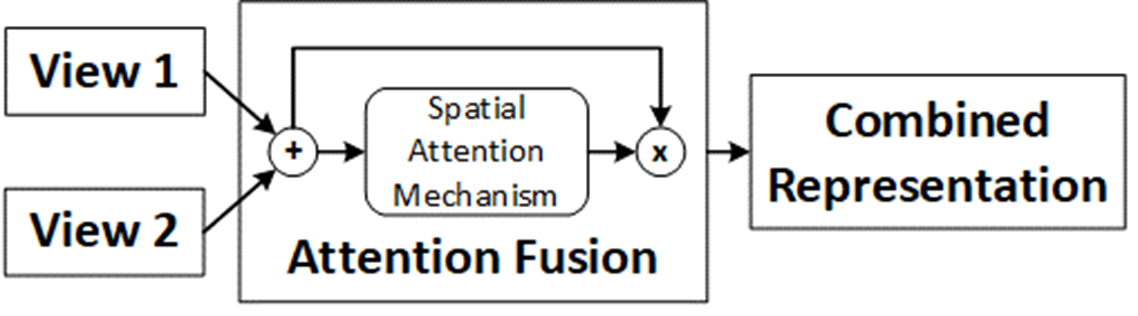}
    \vspace{-8pt}
    \caption{Attention Fusion Basic Architecture}
    \label{attnFusion}
\vspace{-5pt}
\end{figure}

Unfortunately, attention mechanisms can suffer from overfitting during training, attending to specific areas of the input (i.e., generating irrelevant contexts) \cite{anand2021adversarial}. This may result in poor performance in adversarial robustness when attended areas are perturbed in an adversarial attack.

\subsection{Highway Layer}

The shortcomings of attention-based fusion mechanisms can be fixed using gates to filter irrelevant contexts \cite{anand2021adversarial}. One such renowned gating method is the Highway Layer (Figure \ref{hwLayer})\cite{srivastava2015training}. 

\begin{figure}[!h]
\centering
\vspace{-10pt}
    \includegraphics[width=.5\columnwidth]{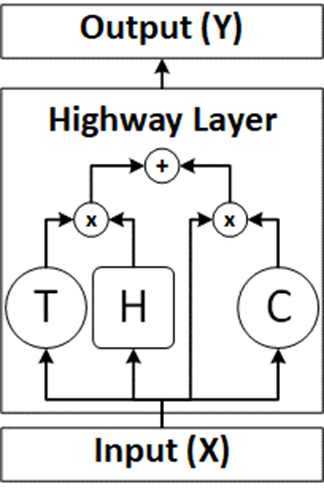}
    \vspace{-10pt}
    \caption{Highway Layer Architecture}
    \label{hwLayer}
    \vspace{-9pt}
\end{figure}

The Highway Layer operates using two gates for information control: the Transformation \((T)\) and the Carry \((C)\) Gates. It is defined using the following equation: 

\[Y = T(X,W_T) * H(X,W_H)  + X * C(X)\]

The Transformation Gate \(T\) determines how much of input \(X\) is subjected to nonlinear transformation \(H\) while the Carry Gate \(C\) determines how much of input \(X\) is allowed through the layer as is.

Highway Layer-based Networks have yielded significant performance in classification tasks, showing their ability to effectively control information flow \cite{li2022deep} \cite{guo2021identifying}. However, their use as fusion mechanisms for MV learning is understudied. 

\section{Research Gaps and Questions}

Based on our literature review, two research gaps are identified. First, within the MV-learning domain, while MV learning has been shown to improve malware classification accuracy, its impact on a model's adversarial robustness is unclear. Second, regarding the methodology, while highway layers has shown promising performance in natural language processing tasks, it is unclear how it could be applied in malware analysis to enhance adversarial robustness of malware detectors against AMG. To address the identified gaps, the following research question is posed: 

\begin{itemize}
    \item{How can we leverage MV learning to create enhanced malware detectors that are robust against adversarial variants?} 
\end{itemize}

\begin{figure*}[t]
    \centering
    \vspace{-10pt}
    \includegraphics[width=0.85\textwidth]{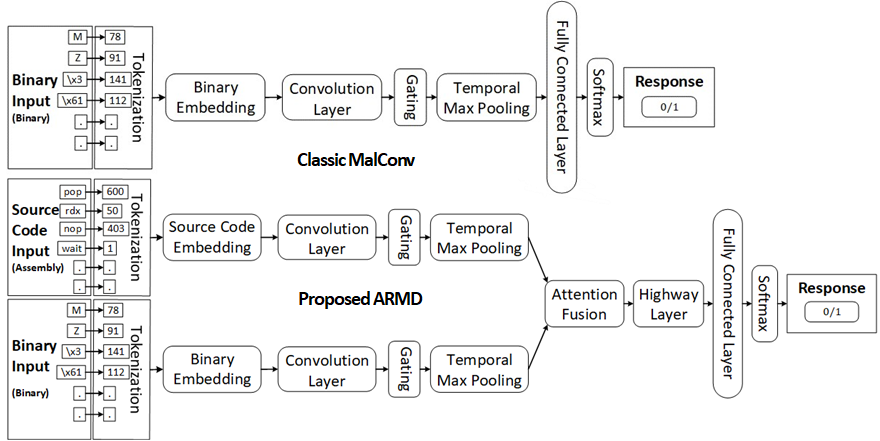}
    \vspace{-10pt}
    \caption{ARMD's Architecture vs. MalConv Architecture}
    \vspace{-9pt}
    \label{armdArch}
\end{figure*}

\begin{figure*}[b]
    \centering
    \vspace{-10pt}
    \includegraphics[width=0.9\textwidth]{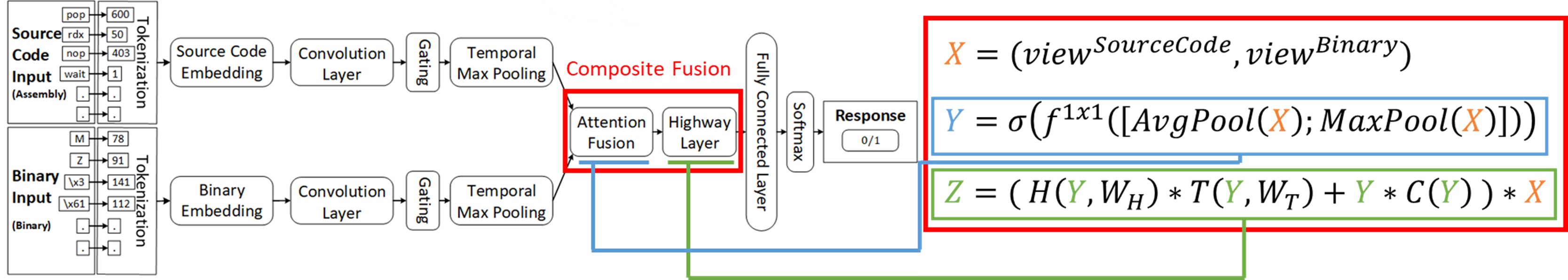}
    \vspace{-10pt}
    \caption{Illustration of ARMD's Novelty (Composite Fusion Mechanism)}
    \vspace{-9pt}
    \label{novelty}
\end{figure*}
\section{Research Design}
Motivated by our research question, we propose an Adversarially Robust Multi-view Malware Defense (ARMD), an MV learning framework to improve the adversarial robustness of malware detectors. 

Following previous AMG studies, we first introduce the threat model that our proposed ARMD will defend against. Next, we introduce the overall architecture of ARMD. Then, we highlight ARMD's novelty, its composite fusion mechanism, in detail. Finally, we present the dataset we use to train and evaluate ARMD.

\subsection{Threat Model}

A threat model is a systematic representation of cyber attacks \cite{carlini2019evaluating}\cite{biggio2013security}. To assess the performance of ARMD against a realistic adversary, our threat model focuses on a black-box setting. Accordingly, three major components of our threat model are:

\begin{itemize}
  \item \textbf{Adversary’s Goal:} Evade ARMD through the generation of adversarial malware samples.
  \item \textbf{Adversary’s Knowledge:}
  \begin{itemize}
      \item Structure and parameters of ARMD are unknown to the attacker.
      \item Attacker does not have access to the confidence score produced by ARMD (fully black-box attack).
  \end{itemize}
  \item \textbf{Adversary's Capability:} The adversary applies functionality-preserving modifications to malware samples.
\end{itemize}

\subsection{ARMD Architecture}\label{ModelFrameworkSection}

To defend against the aforementioned threat model, ARMD extends the MalConv architecture through a separate detector architecture, a composite fusion mechanism, and a softmax decision layer (Figure \ref{armdArch}). 

ARMD's separate detector architecture allows ARMD to learn representations of both views simultaneously. Its composite fusion mechanism uses an attention layer to learn contexts of both representations and a highway layer to filter out irrelevant contexts. Lastly, ARMD's softmax decision Layer makes the final decision based on fused representations, analogous to MalConv's softmax layer. Our intuition here is that through the gating mechanisms of the highway layer, ARMD can filter out irrelevant contexts generated by the attention fusion, thus increasing adversarial robustness. 

\subsection{ARMD Novelty: Composite Fusion Mechanism}
\label{ModelNovelty}

ARMD’s novelty lies in the proposed composite fusion mechanism, highlighted in Figure~\ref{novelty}. The composite fusion mechanism is detailed in the accompanying equations and is shown as follows:
\vspace{-5mm}
\[X = (view^{SourceCode}, view^{Binary})\]
\[Y = \sigma(f^{1x1}([AvgPool(X);MaxPool(X)]))\]
\[Z = (H(Y,W_H) * T(Y,W_T) + Y * C(Y)) * X\]

Within our equations, \(\textit{X}\) denotes the combined representation from both views, connected through a simple concatenation. \(\textit{Y}\) denotes the contexts generated from combined representation throughuy7 the average and max pooling of our spatial attention mechanism. Finally, \(\textit{Z}\) denotes the attention contexts filtered by the highway layer applied back to the original combined representation \({X}\). Our intuition here is that as the highway layer parses the generated attention contexts \(Y\), the transformation gate \((T(Y, W_T ))\) learns to minimize the effects of irrelevant contexts with the nonlinear transformation \((H(Y, W_H ))\) while the carry gate \((C(Y))\) allows relevant contexts through as is. When the filtered contexts are applied back to the combined representation \({X}\), the resulting \(\textit{Z}\) contains the fused representation with only the parts untouched by AMG methods attended to, ready to be passed into ARMD's softmax decision layer for a final classification.

\subsection{Traning \& Validation Testbed} \label{testbedSection}

To train and evaluate ARMD, we collected 33,853 benign and malicious files. For benign files, we collected 13,554 windows system files. For malicious files, we obtained 20,299 malware samples in eight renowned malware categories from VirusTotal, following the example of previous AMG studies \cite{ebrahimi2021binary}\cite{demetrio2021functionality}\cite{hu2021single}\cite{ebrahimi2020binary}\cite{castro2019armed}\cite{chen2019adversarial}\cite{fang2019evading}\cite{rosenberg2019defense}\cite{suciu2019exploring}\cite{anderson2018learning} (Table \ref{trainingTestbedTable}).

\begin{table}[!ht]
\centering
\begin{center}
\vspace{-4mm}
\caption{Dataset used for Model Training}
\vspace{-2mm}
\begin{tabular}{
|m{1.2cm}<{\centering}
|m{3.5cm}<{\centering}
|m{1.75cm}<{\centering}
|m{0.6cm}<{\centering}
|}

\hline

\textbf{Malware Category} &\textbf{Description} &\textbf{Examples} &\textbf{\# of Files}\\

\hline

\textbf{Adware} & Shows unwanted ads and force internet traffic to sites & eldorado, razy, gator & 3,627\\

\hline

\textbf{Backdoor} & Negates normal authentications for host access & lunam, rahack, symmi & 2,160\\

\hline

\textbf{Botnet} & A network of bots connected through the internet & virut, salicode, sality & 1,765\\

\hline

\textbf{Dropper} & Secretly installs other malwares on the host & dinwod, gepys, doboc & 668\\

\hline

\textbf{\makecell{Ransom-\\ware}} & Encrypts data and files, restricting access until decrypted by malware authors & vtflooder, msil, bitman & 1,422\\

\hline

\textbf{Rootkit} & Grants admin privilege to malware author & onjar, dqqd, shipup & 481\\

\hline

\textbf{Spyware} & Allows malware authors to covertly steal personal data & mikey, qqpass, scar & 1,172\\

\hline

\textbf{Virus} & Corrupts files on the host system & nimda, shodi, hematite & 1,756\\

\hline

\textbf{MISC} & Samples that couldn't be organized into previous categories & N/A & 7,248\\

\hline

\textbf{Total} & - & - & \textbf{20,299}\\

\hline
\end{tabular}
\label{trainingTestbedTable}
\vspace{-4mm}
\end{center}
\end{table}

After combining and shuffling together both the benign and malicious files, we split our 33,853 file dataset into a 27,083 file training set and a 6,770 file validation set. We then used our training set to train ARMD while reserving our validation set for later evaluation. 

\section{Evaluation}

\subsection{Experiment Overview}
We conducted three experiments to evaluate ARMD’s performance (Table \ref{experimentOverview}). Experiment \#1 evaluates ARMD and each benchmark method in malware detection. Experiment \#2 evaluates ARMD's and each benchmark method's adversarial robustness when under attack from a state-of-the-art AMG method. Finally, Experiment \#3 evaluates the impact of ARMD's composite fusion mechanism on performance and adversarial robustness through an ablation analysis of ARMD with alternate fusion mechanisms.

\begin{table}[!h]
\centering
\vspace{-4mm}
\begin{center}
\caption{Overview of Experiment}
\vspace{-4mm}
\begin{tabular}{
|m{1.2cm}<{\centering}
|m{2.5cm}<{\centering}
|m{2.3cm}<{\centering}
|m{1.2cm}<{\centering}|}
\hline
\textbf{Experi-ment} & \textbf{Goal} & \textbf{Data Used} & \textbf{Evaluation Metric}\\

\hline

Experiment \#1 & Evaluate effect of MV learning on malware classification performance & Validation set from section \ref{testbedSection}, unseen in the training process & Recall, F1\\

\hline

Experiment \#2 & Evaluate effect of MV learning on adversarial robustness & 4 prominent categories of malware samples unseen in Experiment \#1 and the training process & Evasion Rate\\

\hline

Experiment \#3 & Ablation study on the effect of different fusion mechanisms on malware classification performance and adversarial robustness & Data from previous 2 experiments is used to evaluate malware classification and adversarial robustness, respectively. & F1 \& Evasion Rate\\

\hline
\end{tabular}
\label{experimentOverview}
\vspace{-10mm}
\end{center}
\end{table}

The data and experiment parameters are different for each experiment. 

\subsection{Experiment \#1 – Malware Detection Performance}

To evaluate ARMD’s MV learning’s effect on malware classification, several renowned single-view models were selected as benchmarks (Table \ref{ex1overview}).

\begin{table}[!ht]
\centering
\vspace{-4mm}
\begin{center}
\caption{Overview of Experiment Benchmark Methods}
\vspace{-2mm}
\begin{tabular}{
|m{1.25cm}<{\centering}
|m{2.5cm}<{\centering}
|m{1.75cm}<{\centering}
|m{1.5cm}<{\centering}|}
\hline
\textbf{Method} & \textbf{Description} & \textbf{Owner} & \textbf{Reference(s)}\\

\hline

MalConv & The original MalConv described from literature & Laboratory for Physical Sciences & Raff et al., 2018\\

\hline

NonNeg & Version of MalConv trained with non-negative weight constraints for boosted performance & Laboratory for Physical Sciences & Fleshman et al., 2019\\

\hline

ConvNet & DL-based CNN detector with multiple convolution layers for extensive feature extraction and learning  & Avast & Krcal et al., 2018\\

\hline


\end{tabular}
\label{ex1overview}
\vspace{-4mm}
\end{center}
\end{table}

After training ARMD, the remaining 6,770 file dataset is used as the validation set. Each benchmark is also evaluated using the validation set. For each model, accuracy, precision, recall, and F1 are recorded (Table \ref{ex1result}). 


\begin{table}[!ht] 
\centering
\setlength{\abovecaptionskip}{0pt}
\setlength{\belowcaptionskip}{-10mm}
\begin{center}
\vspace{-4mm}
\caption{Experiment \#1 Results}
\vspace{-2mm}
\begin{tabular}{
|m{2cm}<{\centering}
|m{1cm}<{\centering}
|m{1cm}<{\centering}
|m{1cm}<{\centering}
|m{1.7cm}<{\centering}|}

\hline

\textbf{Metric} & \textbf{MalConv} & \textbf{Nonneg} & \textbf{ConvNet} & \textbf{ARMD (Our Method)}\\

\hline

\textbf{Accuracy (\%)} & 87.93\% & 55.23\% & 73.71\% & \textbf{97.10}\%\\

\hline

\textbf{Precision (\%)} & \textbf{98.89}\% & 98.25\% & 91.72\% & 96.81\%\\

\hline

\textbf{Recall (\%)} & 82.75\% & 33.19\% & 66.42\% & \textbf{98.89}\%\\

\hline

\textbf{F1 (\%)} & 90.10\% & 49.62\% & 77.05\% & \textbf{97.84}\%\\

\hline

\end{tabular}
\vspace{1mm}

\centering \textbf{Note:} P-Values are significant at 0.05.

\label{ex1result}
\vspace{-5mm}
\end{center}
\end{table}

As seen in Table~\ref{ex1result}, our results show a significant increase in ARMD’s performance compared to single-view models, which is consistent with previous research. The implementation of MV architecture in a malware detector improves said detector's malware detection performance. As such, our implementation of MV architecture for a more adversarial robust malware detector will not sacrifice the detector's malware detection performance. 

\setcounter{table}{8}
\begin{table*} [b!] 
\centering
\setlength{\abovecaptionskip}{0pt}
\setlength{\belowcaptionskip}{-10mm}
\begin{center}
\vspace{-4mm}
\caption{Ablation Study Results}
\vspace{-2mm}
\begin{tabular}{
|m{2cm}<{\centering}
|m{2cm}<{\centering}|
|m{2cm}<{\centering}
|m{2cm}<{\centering}
|m{2cm}<{\centering}
|m{2cm}<{\centering}
|m{3cm}<{\centering}|}

\hline

\textbf{Sub-Experiment} & \textbf{Metric / Category} & \textbf{Concatenation} & \textbf{Attention} & \textbf{Highway} & \textbf{Highway + Attention} & \textbf{Attention + Highway (Our method)} \\

\hline
\multirow{3}{*}{\makecell[c]{
\vspace{-1mm}\\\textbf{Malware}\\\textbf{Detection}\\\textbf{Performance}}} & Accuracy (\%) & 97.80\% & \textbf{97.82\%} & 97.81\% & 97.27\% & 97.14\% \\

\cline{2-7}

 & Precision (\%) & 97.94\% & 98.05\% & 97.34\% & \textbf{98.27\%} & 96.49\% \\

\cline{2-7}

 & Recall (\%) & 98.76\% & 98.69\% & \textbf{99.43\%} & 97.61\% & 99.30\% \\

\cline{2-7}

 & F1 (\%) & 98.35\% & 98.37\% & \textbf{98.37\%} & 97.94\% & 97.87\% \\

\cline{2-7}
\cline{2-7}
\hline
\cline{2-7}
\cline{2-7}


\multirow{3}{*}{\makecell{
\vspace{1mm}\\\textbf{Adversarial}\\\textbf{Robustness}\\\textbf{Evaluation}}} & Botnet & 5.32\% & 70.34\% & 10.46\% & 10.27\% &\textbf{5.13\%} \\

\cline{2-7}

 & Ransomware & 15.67\% & 18.67\% & 18.00\% & 17.78\% & \textbf{15.44\%} \\

\cline{2-7}

 & Rootkit & 13.21\% & 20.75\% & 11.32\% & 18.87\% & \textbf{11.32\%} \\

\cline{2-7}

 & Spyware & 7.82\% & 13.30\% & 8.76\% & 10.33\% & \textbf{7.51\%} \\

\cline{2-7}

 & Virus & 10.17\% & 36.42\% & \textbf{8.80\%} & 13.05\% & 10.02\% \\

\cline{2-7}

 & Total & 10.55\% & 31.47\% & 12.13\% & 13.54\% & \textbf{10.30\%} \\

\hline

\end{tabular}
\vspace{1mm}

\centering \textbf{Note:} P-Values are significant at 0.05.

\label{ex3Result}
\vspace{-5mm}
\end{center}
\end{table*}

\subsection{Experiment \#2 – Adversarial Robustness}

To evaluate ARMD’s MV learning’s effect on adversarial robustness, we selected a recent, high-performing Reinforcement Learning (RL) method, adapted from Ebrahimi et al. 2021\cite{ebrahimi2021binary}, as our AMG method, which includes perturbation affecting both malware binary and source code contents. We ran the RL AMG method against ARMD and our detector benchmarks. For our data, we selected a new dataset, unseen in Experiment \#1, of four prominent malware categories to run the RL AMG method on (Table \ref{ex2overview}).

\setcounter{table}{5}
\begin{table}[!h]
\centering
\begin{center}
\vspace{-4mm}
\caption{Malware Samples used in Experiment \#2}
\vspace{-4mm}
\begin{tabular}{
|m{1.2cm}<{\centering}
|m{3.7cm}<{\centering}
|m{1.75cm}<{\centering}
|m{0.5cm}<{\centering}
|}
\hline
\textbf{Malware Category} &\textbf{Description} &\textbf{Examples} &\textbf{\# of Files}\\

\hline
\textbf{Botnet} & A network of bots connected through the internet & virut, salicode, sality & 526\\

\hline
\textbf{\makecell{Ransom-\\ware}} & Encrypts data and files, restricting access and usage until decrypted by malware authors & vtflooder, msil, bitman & 900\\

\hline
\textbf{Rootkit} & Grants admin privilege to malware author & onjar, dqqd, shipup & 53\\

\hline
\textbf{Spyware} & Allows malware authors to steal personal information covertly & mikey, qqpass, scar & 640\\

\hline
\textbf{Virus} & Corrupts files on the host system & nimda, shodi, hematite & 659\\

\hline
\textbf{Total} & - & - & \textbf{2,778}\\

\hline

\end{tabular}
\label{ex2overview}
\vspace{-4mm}
\end{center}
\end{table}

Table \ref{ex2results} shows the evasion rate obtained from running the RL AMG method against each model.

\begin{table}[!ht] 
\centering
\setlength{\abovecaptionskip}{0pt}
\setlength{\belowcaptionskip}{-10mm}
\vspace{-4mm}
\begin{center}
\caption{Experiment \#2 Results}
\vspace{2mm}
\begin{tabular}{
|m{1.5cm}<{\centering}
|m{1cm}<{\centering}
|m{1cm}<{\centering}
|m{1cm}<{\centering}
|m{1cm}<{\centering}
|m{1.5cm}<{\centering}|}

\hline

\textbf{Category} & \textbf{MalConv} & \textbf{Nonneg} & \textbf{ConvNet} & \textbf{ARMD (Our Method)}\\

\hline

\textbf{Botnet} & 68.29\% & 70.72\% & 61.60\% & \textbf{5.13\%}\\

\hline

\textbf{\makecell{Ransom-\\ware}} & 64.28\% & 41.78\% & \textbf{14.89\%} & 15.44\%\\

\hline

\textbf{Rootkit} & 69.23\% & 88.68\% & 56.60\% & \textbf{11.32\%}\\

\hline

\textbf{Spyware} & 80.00\% & 36.93\% & 27.23\% & \textbf{7.51\%}\\

\hline

\textbf{Virus} & 85.71\% & 96.21\% & 67.83\% & \textbf{10.02\%}\\

\hline

\textbf{Total} & 73.84\% & 59.96\% & 39.94\% & \textbf{10.30\%}\\

\hline

\end{tabular}
\vspace{1mm}

\centering \textbf{Note:} P-Values are significant at 0.05.

\label{ex2results}
\vspace{-5mm}
\end{center}
\end{table}

From our Experiment \#2 results, we observe ARMD outperforming every other single-view benchmark detector in almost every category when detecting adversarial samples generated by the RL AMG method. This supports our expectation that a MV malware detector is more adversarial robust than a single-view detector. This is due to the MV detector forcing adversaries to conduct AMG attacks on both views simultaneously. While the chosen RL AMG method can conduct MV attacks, its simultaneous effect on both views is weaker than its effect on a single view. As such, ARMD was able to defend against the selected RL AMG method more so than our single-view benchmarks.

\subsection{Experiment \#3 – Ablation Analysis}

To evaluate how different fusion mechanisms affect adversarial robustness, ARMD’s fusion mechanism was modified with several alternative mechanisms (Table \ref{ex3overview}).

\begin{table}[!ht]
\centering
\vspace{-4mm}
\begin{center}
\caption{Overview of Alternative Fusion Mechanisms}
\vspace{-4mm}
\begin{tabular}{
|m{1.5cm}<{\centering}
|m{1.25cm}<{\centering}
|m{4.8cm}<{\centering}|}

\hline

\textbf{Fusion Mechanism} & \textbf{Reference} & \textbf{Description}\\

\hline

Concatenation & Gao et al., 2020 & Source code and binary representations are concatenated\\

\hline

Attention & Jetley et al., 2018 & An attention mechanism is used to determine the importance of different areas of the input.\\

\hline

Highway & Srivstava et al., 2015 & Several gates are employed to control the flow of information from the input layer to the convolution layers\\

\hline

Highway + Attention & - & Composite fusion with a Highway Layer fusion followed by an attention layer to pass along information.\\

\hline

\textbf{Attention + Highway (Our method)} & - & Composite fusion with an attention layer fusion followed by a highway layer to pass along information.\\

\hline

\end{tabular}
\label{ex3overview}
\vspace{-4mm}
\end{center}
\end{table}

Experiment \#3 consists of two parts, malware classification performance and adversarial robustness evaluation. Data from Experiment \#1 is used for malware detection performance, with model performance evaluated using F1 score. Data from Experiment \#2 is used for adversarial robustness evaluation, with model performance evaluated using evasion rate. After running the RL AMG method against each fusion mechanism, their respective F1 scores and evasion rates are recorded (Table \ref{ex3Result}).

Our results show that our proposed fusion mechanism performs comparably with other alternatives in malware detection while outperforming other alternatives in adversarial robustness. 

\subsection{Experiment Result Summary}
We derived three important observations from our experiment results. (1) Experiment \#1 demonstrates that implementing MV learning into malware detectors increases a detector’s malware detection performance. (2) Experiment \#2 indicates MV learning’s ability to improve a model’s adversarial robustness. (3) Experiment \#3 shows ARMD’s specific fusion mechanism (Attention + Highway) as the most adversarially robust fusion mechanism with minimal effect on the detection performance. These observations suggest that as ARMD operates, its highway layer filters out the irrelevant contexts generated by its attention layer, thus improving ARMD's adversarial robustness and supporting our initial intuitions.








\section{Conclusion and Future Directions}

Deep Learning (DL)-based malware detectors are susceptible to Adversarial Malware Generation (AMG) techniques. Based on the nature of these adversarial modifications, these AMG methods are often limited to operating on the binary view of a malware sample and do not account for alternative views representing the sample. Recognizing this gap as an opportunity for improving defense, in this study, we developed a novel detector architecture that considers two views of the malware, combining each representation through a novel composite fusion mechanism. Our results show that ARMD is significantly more adversarially robust against AMG attacks without sacrificing too much malware detection performance. This highlights the viability of leveraging the single-view nature of AMG techniques to boost a detector’s adversarial robustness through multi-view learning. Considering our results, we recommend malware detector developers incorporate multi-view capabilities into their DL-based detectors to enhance adversarial robustness.

Based on our results, we identify three promising directions for future research. First, we envision expanding the number of views ARMD can consider (e.g., including a malware’s API calls) so that ARMD can capture behavioral data about a malware file to render ARMD more invulnerable to single-view AMG attacks. Second, we believe evaluating ARMD on polymorphic AMG techniques, such as file encryption through the open-source tool darkarmour, can show how ARMD handles massive, file-wide perturbations. Lastly, implementing ARMD in a live-fire cybersecurity environment where real-world hackers can try to evade ARMD with their custom-made adversarial samples will yield interesting results on how well our model can perform in a real-world setting.

\section*{Acknowledgment}
We would like to sincerely thank VirusTotal for providing the malware dataset and granting access to the corresponding APIs for functionality assessment.

This material is based upon work supported by the National Science Foundation (NSF) under Secure and Trustworthy Cyberspace (grant No. 1936370), Cybersecurity Innovation for Cyber Infrastructure (grant No. 1917117), and Cybersecurity Scholarship-for-Service (grant No. 1921485).

\bibliographystyle{IEEEtran}
\bibliography{main}

\vspace{12pt}
\color{red}

\end{document}